\documentstyle[12pt,pictex]{article}
\begin{document}

\centerline{\Large\bf Effective Potentials and Symmetry Restoration }
\centerline{\Large\bf in the Chiral Abelian Higgs Model}

\vskip 10mm
\centerline{Jens O. Andersen}
\centerline{\it Institute of Physics}
\centerline{\it University of Oslo}
\centerline{\it P.O. BOX 1048, Blindern}
\centerline{\it N-0316 Oslo, Norway}

\vspace{10mm}
\begin{center}
{\bf Abstract}
\end{center}
{\footnotesize The chiral Abelian Higgs model is studied at finite
temperature. By
integrating out the heavy modes, we make a three-dimensional effective theory
for the static modes.
It is demonstrated that
the plasma masses are correctly reproduced to leading order in $m^{2}/T^{2}$.
The
effective potential for the composite operator $\phi^{\dagger}\phi$
is calculated at one loop for the resulting three-dimensional theory
and it is shown that the result is gauge parameter invariant. The numerical
investigation of the potential reveals that the symmetry is restored via
a first order phase transition.
Comparison with the ordinary ring improved potential is made and it is found
that the barrier height at $T_{c}$ is somewhat higher. }
\small
\normalsize
\section{Introduction}
Quantum field theories at finite temperature has received considerable
attention, since the work of Dolan and Jackiw twenty years ago \cite{dolan}.
It was first observed by Kirznitz and Linde \cite{kirz} that symmetries
which are
spontaneously broken at zero temperature are normally restored
at high temperatures.
Recently, there has been much interest in the electroweak phase transition,
mainly because of its role in the generation of the baryon asymmetry of the
universe \cite{cohen}.
Important aspects are the order and strength of the phase transition as a
function of the Higgs boson mass, and the calculation of nucleation rates.

Several approaches have been used in the the study of the electroweak phase
transition. This includes the use of three-dimensional effective theories
[4,5],
the $\epsilon$-expansion \cite{arnold3} and ring improved effective
potentials \cite{car}.
Recently, Buchm\"uller {\it et al.} have proposed an effective potential for
the operator $\sigma =\phi^{\dagger}\phi$ and have applied it to the
Abelian Higgs model and SU(2) [8,9]. It has been claimed that this new
effective potential is gauge invariant, which is a desireable property.
The problems of gauge invariance and gauge fixing dependence are important
and must be taken seriously.
Physical quantities such as the critical temperature must be gauge invariant.
Normally, the effective potential is gauge fixing independent at the minimum,
that is
for the value of the field that is a solution of the effective field
equations (``on shell'').
However, the potential should also be gauge invariant away from the minimum
(``off shell'') since the form of the potential (e.g. barrier height)
determines the dynamics of the phase transition.

In the present work we consider the chiral Abelian Higgs model with a
quartic self interaction in 3+1 dimensions. This model has previously been
studied by Arnold and Espinosa \cite{arnold1}, using resummation techniques.
We shall use a different approach
by making a three-dimensional effective theory for the static modes
by integrating out the heavy modes. This method has been extensively used
in finite temperature field theory in recent years. Applications
to U(1) and SU(2) gauge theories are found in refs. [4,5] and we recommend
them for further details.

The decoupling of the non-static (heavy) modes from the high temperature
dynamics was proposed a long time ago and has been studied in detail in various
theories, e.g. QCD and QED \cite{jour}. By doing this dimensional reduction
[11-13] the infrared behaviour of the static modes improves due to the
induced thermal masses. Secondly, this
approach induces non-linear interaction between the static modes.

The dimensional reduction is carried out in section two and it is explicitly
demonstrated that one obtains the same thermal masses to
leading
order in $m^{2}/T^{2}$ as those found by solving the Schwinger-Dyson equation
in the
full four-dimensional theory \cite{arnold1}.
In section three we compute the effective potential for the composite
operator
$\sigma =\phi^{\dagger}\phi$ to one loop order. The calculations are
carried out in the
$R_{\xi}$ gauge and it is shown that the effective potential is
independent of the gauge parameter at one loop. Although this does
not imply gauge invariance of the effective action, this independence is
of course a nice feature.
The potential is investigated numerically at finite temperature and is compared
with  the ordinary ring improved effective potential. It is found that the
symmetry is restored at high temperature via a first order phase transition.
Finally, we summarize and make some comments on further developments
in section four.
\section{The Three-dimensional Effective Theory}
{\it The Abelian Higgs model.}$\,\,$
Let us first consider the Abelian Higgs model without fermions.
The Euclidean Lagrangian reads:
\begin{eqnarray}
{\cal L}&=&\frac{1}{4}F_{\mu\nu}F_{\mu\nu}+\frac{1}{2}
(D_{\mu}\Phi)^{\dagger}(D_{\mu}\Phi)-\frac{1}{2}c^{2}\Phi^{\dagger}\Phi
+\frac{\lambda}{4}(\Phi^{\dagger}\Phi)^{2}.
\end{eqnarray}
plus gauge fixing terms. Here $D_{\mu}=\partial_{\mu}+ieA_{\mu}$ is the
covariant
derivative. The corresponding action is then
\begin{equation}
S=\int_{0}^{\beta}d\tau\int {\cal L}\,d^{3}{\bf x}.
\end{equation}
At finite temperature we expand the fields as
\begin{eqnarray}
A_{i}({\bf x},\tau)&=&\beta^{-\frac{1}{2}}\Big [\,A_{i}({\bf x})
+\sum_{n\neq 0}a_{i,n}({\bf x})e^{2\pi in\tau/\beta}\,\Big ],
\hspace{1cm} i=1,2,3\\
A_{\tau}({\bf x},\tau)&=&\beta^{-\frac{1}{2}}\Big [\,\rho ({\bf x})
+\sum_{n\neq 0}a_{\tau,n}({\bf x})e^{2\pi in\tau/\beta}\,\Big ]\\
\Phi ({\bf x},\tau)&=&\beta^{-\frac{1}{2}}\Big [\,\phi_{0}({\bf x})
+\sum_{n\neq 0}\phi_{n}({\bf x})e^{2\pi in\tau/\beta}\,\Big ].
\end{eqnarray}
The calculations will be carried out in the thermal static gauge
\cite{jakovac}:
\begin{equation}
a_{\tau,n}({\bf x})=0,\hspace{1cm} \forall \,\,n. \\
\end{equation}
We integrate over $\tau$ and exploit the orthonormality of the modes.
There will now be terms in the action which involve only the static modes
\begin{eqnarray} \nonumber
\label{static}
S^{(0)}&=&\int \Big [\,\frac{1}{4}F_{ij}F_{ij}+\frac{1}{2}
(\partial_{i}\rho)^{2} +\frac{1}{2}(\partial_{i}\phi_{0})^{\dagger}
(\partial_{i}\phi_{0})-\frac{1}{2}c^{2}\phi_{0}^{\dagger}\phi_{0}
+\frac{\lambda T}{4}(\phi_{0}^{\dagger}\phi_{0})^{2}\\
&&+\frac{e^{2}T}{2}(A_{i}^{2}+\rho^{2})\phi_{0}^{\dagger}\phi_{0}
+\frac{ieA_{i}T^{\frac{1}{2}}}{2}(\phi_{0}\partial_{i}\phi_{0}^{\dagger}
-\phi_{0}^{\dagger}\partial_{i}\phi_{0})\,\Big ]\,d^{3}{\bf x}.
\end{eqnarray}
The terms that are quadratic in non-static modes are
\begin{eqnarray}\nonumber
\label{kvad}
S^{(2)}&=&\sum_{n\neq 0}\int\Big [\frac{1}{2}(\partial_{i}
a_{j,n})^{\dagger}(\partial_{i}a_{j,n})-\frac{1}{2}
(\partial_{i}a_{i,n})^{\dagger}(\partial_{i}a_{i,n})
+\frac{1}{2}(2\pi nT)^{2}a_{i,n}^{\dagger}a_{i,n}\\
&&+\frac{1}{2}(2\pi nT)^{2}\phi_{n}^{\dagger}\phi_{n}
+\frac{1}{2}(\partial_{i}\phi_{n})^{\dagger}(\partial_{i}\phi_{n})
-\frac{1}{2}c^{2}\phi_{n}^{\dagger}\phi_{n}\,\Big ]\,d^{3}{\bf x},
\end{eqnarray}
and finally there are  terms representing the interactions between
the static and
the non-static modes.
These terms generate the effective thermal masses of
the zero modes $\phi_{0}$ and $\rho$
\begin{eqnarray}
\label{eq:int}\nonumber
S^{(2)}_{int}&=&\sum_{n\neq 0}\int\Big [\frac{1}{2}e^{2}Ta_{i,n}
^{\dagger}a_{i,n}(\phi_{0}^{\dagger}\phi_{0})
+e\rho 2\pi nT^{\frac{3}{2}}\phi_{n}^{\dagger}\phi_{n}
+\frac{\lambda T}{4}4\phi_{0}^{\dagger}\phi_{0}\phi_{n}^{\dagger}\phi_{n}
\\
&&+\frac{ieT^{\frac{1}{2}}}{2}a_{i,n}(\phi_{0}\partial_{i}\phi_{n}^{\dagger}
+\phi_{-n}\partial_{i}\phi_{0}^{\dagger}-\phi_{0}^{\dagger}\partial_{i}
\phi_{-n}-\phi_{n}^{\dagger}\partial_{i}\phi_{0})
+\frac{1}{2}e^{2}\rho^{2}T(\phi_{n}^{\dagger}\phi_{n})\Big ]\,d^{3}{\bf x}.
\end{eqnarray}
We should make the remark that we in eq. (\ref{eq:int}) have set
$A_{i}({\bf x})=0$
since these terms only affect the kinetic part of the effective theory
(see refs. [15,16]).
The omission of these terms then correspond to the neglect of wave function
renormalization and also some finite higher order corrections to the
interaction between
$\rho$ and $\phi_{0}$ and to the scalar potential of the $\rho$-field.
These corrections are of order $e^{4}$ and should hence not be included
since we
calculate consistently to order $e^{2}$. The
fact that the spatial part of $A_{\mu}(x)$ remains massless and therefore acts
as the gauge field in the dimensionally reduced theory could also have been
predicted on general grounds by considering Ward identities in the high
temperature
limit [13,18]. \\ \\
Introducing the two real fields by $\phi_{n}=\phi_{1,n}+i\phi_{2,n}$, we
define the propagators
for the fields $a_{i,n}, \phi_{1,n}$ and $\phi_{2,n}$ by:
\begin{eqnarray}
\Big [\,[-\nabla^{2} +(2\pi nT)^{2}]\delta_{ij}+\partial_{i}
\partial_{j}\,\Big ]D_{jk,n}({\bf x},{\bf x}^{\prime})&=&\delta_{ik}\delta
({\bf x},{\bf x}^{\prime}) \\
\Big [-\nabla^{2}-c^{2}+(2\pi nT)^{2}\Big ]\Delta_{i,n} ({\bf x}
,{\bf x}^{\prime})&=&\delta ({\bf x},{\bf x}^{\prime}),\hspace{1cm}i=1,2.
\end{eqnarray}
We may write the effective action for the zero modes as
\begin{equation}
S_{eff}=S^{(0)}+S^{(2)}+\Delta S
\end{equation}
where
\begin{equation}
\label{eq:eff}
\Delta S=\langle S^{(2)}_{int}\rangle -\frac{1}{2}\langle
(S^{(2)}_{int})^{2}\rangle +...,
\end{equation}
and $S^{(2)}$ is the quadratic contribution from eq. (\ref{kvad}).
The second term in eq. (\ref{eq:eff}) is necessary
in order to calculate consistently to order $\lambda$ and $e^{2}$.
Using
the propagators we find the different contributions to the effective theory.
The contribution from the heavy scalar modes to the static scalar mode
is easily found:
\begin{eqnarray}\nonumber
\label{eq:skalar}
\Delta S_{scalar}&=&\sum_{n\neq 0,\,i}\int\lambda T\phi_{0}^{\dagger}
\phi_{0}\Delta_{i,n}({\bf x},{\bf x})\,d^{3}{\bf x} \\ \nonumber
&=&\sum_{n\neq 0}\int \Big [2\lambda T\phi_{0}^{\dagger}\phi_{0}\int
\frac{d^{3}p}{(2\pi)^{3}}\frac{1}{p^{2}-c^{2}+(2\pi nT)^{2}}\Big ]
\,d^{3}{\bf x} \\ \nonumber
&=&\int \Big [\frac{2\lambda T\phi_{0}^{\dagger}\phi_{0}}{2\pi^{2}}
\int_{0}^{\infty}\frac{pdp}{(\exp\beta p-1)}+{\cal O}(\frac{c^{2}}{T^{2}})
\Big ]\,d^{3}{\bf x}\\
&=&\int \Big[\lambda \phi_{0}^{\dagger}\phi_{0} \frac{T^{2}}{6}+{\cal O}
(\frac{c^{2}}{T^{2}})\Big ]\,d^{3}{\bf x}.
\end{eqnarray}
In the above equation we have dropped a divergence, which corresponds to a
mass renormalization
(see also ref. \cite{fend}). The corresponding Feynman diagrams are shown in
fig.~\ref{fey1}a.
We would also like to make the remark that
our renormalization procedure differs from that of Jakov\'ac and
Patk\'os \cite{jakovac}. This is partly due to their introduction of the
auxillary fields $\chi$ by a Hubbard-Stratonovich transformation \cite{hub}.

The contributions from the vector particles
are calculated
in a similar way, and after some lengthy algebra we find
\begin{eqnarray}
\label{eq:ds}
\Delta S=\int \Big[\frac{1}{2}\phi_{0}^{\dagger}\phi_{0}(4\lambda +3e^{2})
\frac{T^{2}}{12} +\frac{1}{6}\rho^{2}e^{2}T^{2}\Big ]\,d^{3}{\bf x}.
\end{eqnarray}
The Feynman diagrams are displayed in figs.~\ref{fey1}b - ~\ref{fey1}c and
{}~\ref{fey2}a - ~\ref{fey2}b\footnote{Note
that the contribution from fig.~\ref{fey1}b depends on the external momentum
{\bf k}. We
have made a high temperature expansion and included the dominant $T^{2}$ term.
The {\bf k}$^{2}$ pieces contribute to wave function renormalization.}.
Our result is to leading order in $c^{2}/T^{2}$ in accordance with that of
Arnold and
Espinosa \cite{arnold1}, who used the Schwinger-Dyson equations for the
propagators. The result is also in agreement with that obtained by
Jakov\'ac and Patk\'os \cite{jakovac} to order $\lambda$ and $e^{2}$.\\ \\
The effective theory for the static modes is now obtained:
\begin{eqnarray}\nonumber
\label{eff}
S_{eff}&=&\int\Big [\,\frac{1}{4}F_{ij}F_{ij}+\frac{1}{2}(\partial_{i}
\rho)^{2}+\frac{1}{2}m^{2}_{\rho}\rho^{2}+\frac{1}{2}(D_{\mu}\phi_{0})
^{\dagger}(D_{\mu}\phi_{0})+\frac{1}{2}m^{2}\phi_{0}^{\dagger}\phi_{0}
+\frac{\lambda T}{4}(\phi_{0}^{\dagger}\phi_{0})^{2} \\
&&+\frac{e^{2}T}{2}\rho^{2}\phi_{0}^{\dagger}\phi_{0}\,\Big ]\,d^{3}{\bf x},
\end{eqnarray}
where the thermal masses are given by:
\begin{eqnarray}
\label{eq:thermal}
m^{2}&=&-c^{2}+(4\lambda +3e^{2})\frac{T^{2}}{12}\\
m^{2}_{\rho}&=&\frac{e^{2}T^{2}}{3}.
\end{eqnarray}
Note that we discard $S^{(2)}$ from the effective theory since it is
independent of the static mode. It only gives a temperature
dependent contribution to the effective action and will not affect the
critical temperature.
{}From eq. (\ref{eff}) one observes that the zeroth component of the vector
potential $A_{\mu}({\bf x},\tau)$ plays the role of an extra scalar field
in this effective theory and
that it is coupled non-linearly to the static mode of the scalar
field $\Phi ({\bf x},\tau)$.\\ \\
{\it The chiral Abelian Higgs model.}$\,\,$
Let us next couple Dirac fermions chirally to the Abelian Higgs model:
\begin{equation}
{\cal L}^{\prime}={\cal L}+g\overline{\Psi}\Big [ \Phi^{\dagger}
(\frac{1-\gamma_{5}}{2})+\Phi (\frac{1+\gamma_{5}}{2})\Big ]\Psi
+\overline{\Psi}\Big [\gamma^{\mu}\partial_{\mu}-ie\gamma^{\mu}A_{\mu}
(\frac{1-\gamma_{5}}{2})\Big ]\Psi.
\end{equation}
At finite temperature we expand the fermionic field as
\begin{equation}
\Psi ({\bf x},\tau)=\beta^{-\frac{1}{2}}\sum_{n}\psi_{n}({\bf x})
e^{\pi i(2n+1)/\beta}\,\,.
\end{equation}
The fermions are antiperiodic in time, which implies that they do not
contribute to $S^{(0)}$ in eq. (\ref{static}). We then write
\begin{eqnarray}
S^{(2)\prime}&=&S^{(2)}+\sum_{n}\int \overline{\psi}_{n}\gamma^{\mu}
\partial_{\mu}\psi_{n}\,d^{3}{\bf x}  \\ \nonumber
S^{(2)\prime}_{int}&=&S^{(2)}_{int}+\sum_{n}\int \Big [gT^{\frac{1}{2}}
\overline{\psi}_{n}\phi_{0}^{\dagger}(\frac{1-\gamma_{5}}{2})\psi_{n}
+gT^{\frac{1}{2}}\overline{\psi}_{n}\phi_{0}(\frac{1+\gamma_{5}}{2})
\psi_{n} \\
&&-ie\rho T^{\frac{1}{2}}\overline{\psi}_{n}\gamma^{0}
(\frac{1-\gamma_{5}}{2})\psi_{n}\Big ]\,d^{3}{\bf x}
\end{eqnarray}
where $\gamma^{\mu}\partial_{\mu}$ now means $\gamma^{i}\partial_{i}
+\gamma^{0}(2n+1)\pi T$. The fermion propagators are defined by
\begin{equation}
( \gamma^{\mu}\partial_{\mu})S_{F,n}({\bf x},{\bf x}^{\prime})
=\delta ({\bf x},{\bf x}^{\prime}),
\end{equation}
We can now compute the fermion contribution to the effective action $S_{eff}$
for the static modes.
The calculations are carried out consistently to order $g^{2}$ applying the
same techniques as previously. After some
manipulations one finds that the fermion contribution to
$\langle S^{(2)\prime}_{int}\rangle $
vanishes identically due to the properties of the gamma matrices.
Thus one is left
with the correction:
\begin{eqnarray}\nonumber
\label{eq:ds2}
\Delta S_{fermion}&=&\sum_{n}Tr\int\Big [ \frac{1}{2}g^{2}T\phi_{0}
^{\dagger}\phi_{0}\Delta_{F,n}({\bf x},{\bf x}^{\prime})
\Big (\frac{1-\gamma_{5}}{2} \Big )\Delta_{F,n}({\bf x},{\bf x}^{\prime})
\Big (\frac{1+\gamma_{5}}{2} \Big )\Big]\,d^{3}{\bf x}\,
d^{3}{\bf x}^{\prime}\\ \nonumber
&&+\sum_{n}Tr\int\Big [ \frac{1}{2}e^{2}T\rho^{2}\Delta_{F,n}({\bf x},
{\bf x}^{\prime})\gamma^{0}\Big (\frac{1-\gamma_{5}}{2} \Big )
\Delta_{F,n}({\bf x},{\bf x}^{\prime})\gamma^{0}\Big (\frac{1-\gamma_{5}}{2}
\Big )\Big]\,d^{3}{\bf x}\,d^{3}{\bf x}^{\prime}\\
&=&\int\Big [\frac{1}{24}\phi_{0}^{\dagger}\phi_{0}g^{2}T^{2}+\frac{1}{12}
\rho^{2}e^{2}T^{2}\Big ]\,d^{3}{\bf x}.
\end{eqnarray}
The diagrams are shown in figs.~\ref{fey1}d and~\ref{fey2}c.
Again, we have only kept the dominant temperature contribution.
Using eqs. (\ref{eq:ds}) and (\ref{eq:ds2}) one obtains the following
effective masses
\begin{eqnarray} \nonumber
\label{eq:thermal2}
m^{2}&=&-c^{2}+(4\lambda +3e^{2}+g^{2})\frac{T^{2}}{12}\\
m^{2}_{\rho}&=&\frac{e^{2}T^{2}}{2}.
\end{eqnarray}
Our result is again in agreement with that of Arnold and
Espinosa \cite{arnold1}.
\section{The One-Loop Effective Potential}
{\it The one-loop effective potential.}$\,\,$
With the effective three-dimensional theory at hand, we now calculate the
effective potential for the composite field
$\sigma =\phi^{\dagger}\phi$ [8,9] in the one-loop approximation. The
calculations for the Abelian Higgs model and the chiral Abelian Higgs model
are identical; the only difference is that different thermal masses enter
into the final result.
(In the following we drop the subscript on the scalar field and hence write
$\phi$ instead of $\phi_{0}$).
In order to do so we compute the free energy in the presence of a constant
external source $J$:
\begin{equation}
e^{-\Omega W(J)}=\int {\cal D}A_{i}{\cal D}\phi^{\dagger}{\cal D}\phi{\cal D}
\rho e^{-S_{eff}(A_{i},\phi^{\dagger},\phi,\rho ) -\int \phi^{\dagger}
\phi J\,d^{3}{\bf x} }\,,
\end{equation}
where $\Omega$ is the three-dimensional volume.
The composite field $\sigma$ is defined through the relation:
\begin{eqnarray}
\label{invert}
\frac{\delta W(J)}{\delta J}&=&\sigma .
\end{eqnarray}
The effective potential is then obtained as a Legendre transform in the usual
way:
\begin{eqnarray}
V(\sigma)&=&W(J)-\sigma J.
\end{eqnarray}
The classical potential is
\begin{equation}
V_{0}(\phi^{\dagger}\phi)=\frac{1}{2}(m^{2}+2J)\phi^{\dagger}\phi
+\frac{1}{4}\lambda T(\phi^{\dagger}\phi)^{2}+\frac{1}{2}m^{2}_{\rho}\rho^{2}
+\frac{1}{2}e^{2}T\rho^{2}\phi^{\dagger}\phi .
\end{equation}
The classical equations of motion then read
\begin{equation}
e^{2}T\rho^{2}\phi+(m^{2}+2J+\lambda T\phi^{\dagger}\phi)\phi=0,\hspace{1cm}
e^{2}T\rho\phi^{\dagger}\phi +m^{2}_{\rho}\rho =0,
\end{equation}
and have two solutions:
\begin{eqnarray}
\label{sym}
\overline{\rho} &=&0,\hspace{1cm} \overline{\phi}=\phi_{s}=0 \\
\label{asym}
\overline{\rho} &=&0,\hspace{1cm} \overline{\phi}=\phi_{b}=
[-\frac{1}{\lambda T}(m^{2}+2J)]^{\frac{1}{2}}e^{i\alpha}.
\end{eqnarray}
Here $\alpha$ is a phase.
The solutions (\ref{sym}) and (\ref{asym}) correspond to the global
minimum of the classical action in the presence of the source $J$ for
$m^{2}+2J>0$ and $m^{2}+2J<0$, respectively.\\ \\
The masses of the particles are given by the following expressions
\begin{eqnarray}
m^{2}_{A}&=&e^{2}T|\overline{\phi}|^{2},\hspace{3.5cm}m^{2}_{\phi}
=m^{2}+2J+3\lambda T|\overline{\phi}|^{2}, \\
m^{2}_{\chi}&=&m^{2}+2J+\lambda T|\overline{\phi}|^{2},\hspace{1.5cm}
m_{\rho}^{2}=\frac{e^{2}T^{2}}{2}+e^{2}T|\overline{\phi}|^{2}. \\
\end{eqnarray}
In the symmetric phase the masses are
\begin{equation}
m^{2}_{A}=0,\hspace{1cm}m^{2}_{\phi}=m^{2}+2J,\hspace{1cm}m^{2}_{\chi}=
m^{2}+2J,\hspace{1cm}m_{\rho}^{2}=\frac{e^{2}T^{2}}{2},
\end{equation}
while in the broken phase one obtains
\begin{equation}
m^{2}_{A}=-\frac{e^{2}}{\lambda }(m^{2}+2J),\hspace{0.8cm}m^{2}_{\phi}
=-2(m^{2}+2J),\hspace{0.8cm}m^{2}_{\chi}=0,\hspace{0.8cm}m_{\rho}^{2}
=\frac{e^{2}T^{2}}{2}-\frac{e^{2}}{\lambda}(m^{2}+2J).
\end{equation}
We shall work in the $R_{\xi}$ gauge, where the gauge fixing term is
\begin{equation}
{\cal L}_{GF}=\frac{1}{2\xi}(\partial_{i}A_{i}+\xi eT^{\frac{1}{2}}
\overline{\phi}\phi_{2})^{2},
\end{equation}
where we have used the global O(2) symmetry to make $\overline{\phi}$
purely real and
$\phi_{2}$ is the imaginary part of the quantum field $\phi$.
This gauge is particularly simple since the cross terms between the scalar
field and the gauge field in the effective theory disappear. This makes it
rather easy to calculate
the one loop correction to the classical potential:
\begin{eqnarray}\nonumber
W(J)&=&\frac{1}{2}\int \frac{d^{3}k}{(2\pi)^{3}}\Big[2\log (k^{2}
+m^{2}_{A})+\log (k^{2}+m^{2}_{\phi}) +\log (k^{2}+m^{2}_{\chi}
+\xi m_{A}^{2})\\
&&+\log (k^{2}
+\xi e^{2}T\overline {\phi}^{2})
\label{1l}
+(k^{2}+m_{\rho}^{2}) \Big ].
\end{eqnarray}
The corresponding ghost contribution is found to be \cite{jakovac}
\begin{equation}
S_{ghost}=-\int \frac{d^{3}k}{(2\pi)^{3}}\log (k^{2}+\xi e^{2}T\overline
{\phi}^{2}).
\end{equation}
Using dimensional regularization (see e.g. ref. \cite{ryder}) the above
integrals are easily computed:
\begin{equation}
\int \frac{d^{3}k}{(2\pi)^{3}}\log (k^{2}+M^{2})=-\frac{1}{6\pi}M^{3}.
\end{equation}
The result is perfectly finite after regularization and is independent of the
renormalization scale $\mu$.\\ \\
We see that the terms involving $\xi$ cancel since one of the masses, either
$m_{A}^{2}$ or $m_{\chi}^{2}$, vanishes. The effective
potential is thus
{\it gauge parameter} independent (to one loop order), in contrast
to the ordinary effective potential \cite{dolan}\footnote{Working in Lorentz
gauge, where ${\cal L}_{GF}=\frac{1}{2\alpha}(\partial_{i}A_{i})^{2}$,
one finds that
the masses and hence the effective potential are explicitly dependent on
the gauge parameter. Furthermore, this dependence disappears at the minimum
of the potential, as explained in the introduction.}. Moreover, it can also be
shown that  eqs. (\ref{eq:sym}) and (\ref{eq:asym}) below can be obtained
in Lorentz gauge.

There is some confusion
in the literature about gauge invariance and gauge parameter independence.
We emphasize that gauge parameter independence and gauge invariance
are related issues, but not equivalent. One may have gauge invariance with
respect to gauge transformations of the background fields, but still have
dependence on how one fixes the gauge of the quantum fields. This is
the case when applying the method of mean field gauges.
See ref. \cite{abbott} for details. \\ \\
In the symmetric phase one finds the free energy
\begin{equation}
\label{eq:sym}
W_{s}(J)=-\frac{1}{6\pi}(m^{2}+2J)^{\frac{3}{2}}.
\end{equation}
In  the broken phase we get
\begin{eqnarray}\nonumber
\label{eq:asym}
W_{b}(J)&=&-\frac{1}{4\lambda T}(m^{2}+2J)^{2}-\frac{1}{6\pi}
\Big [-\frac{e^{2}}{\lambda}(m^{2}+2J)\Big ]^{\frac{3}{2}}
-\frac{1}{12\pi}\Big [-2(m^{2}+2J)\Big ]^{\frac{3}{2}}\\
&&-\frac{1}{12\pi}\Big [\frac{e^{2}T^{2}}{2}
-\frac{e^{2}}{\lambda}(m^{2}+2J)\big ]^{\frac{3}{2}}.
\end{eqnarray}
Notice that the contribution from the $\rho$ particles is independent of $J$
in the symmetric phase.
The contribution to the effective
potential is therefore independent of the background field and is discarded.
A similar remark applies to the vector meson in the symmetric phase,
since the mass vanishes.\\ \\
In the broken phase it is sufficient to use the tree level expression
for the free energy in order to calculate $\sigma$ from eq. (\ref{invert})
\cite{wilf1}.
It is now straightforward to derive the results
\begin{equation}
V_{s}(\sigma)=\frac{1}{2}m^{2}\sigma -\frac{10\pi^{2}}{3}\sigma^{3}
\end{equation}
and
\begin{equation}
V_{b}(\sigma)=\frac{1}{2}m^{2}\sigma +\frac{1}{4}\lambda T\sigma^{2}
-\frac{1}{6\pi}e^{3}(\sigma T)^{\frac{3}{2}}-\frac{1}{12\pi}
(2\lambda\sigma T)^{\frac{3}{2}}-\frac{1}{12\pi}(\frac{e^{2}T^{2}}{2}
+e^{2}\sigma T)^{\frac{3}{2}}.
\end{equation}
{}From eq. (\ref{invert}) one finds that the symmetric phase is
represented by $\sigma <0$ while the broken phase is represented by
$\sigma >0$. Thus, we can write our result as
\begin{equation}
V(\sigma)=V_{b}(\sigma)\Theta (\sigma)+V_{s}(\sigma)\Theta (-\sigma).
\end{equation}
Some comments are in order.
Firstly, we note that we have a contribution to the potential from the
zeroth component of the gauge field. This is in turn a consequence of the
non-linear interaction between the fields $\phi$ and $\rho$ that was
a result of the dimensional reduction.
Secondly, the non-analytic terms in the effective
potential are independent of $m^{2}$. This implies that $V(\sigma)$ is
valid also below the barrier temperature (in the sense that it is purely
real), contrary to the ring improved
effective potential (see eq. (\ref{ring}) below). \\ \\
To lowest order in the couplings [5,8] one has
\begin{equation}
V_{4}(\sigma)=TV_{3}(\frac{\sigma}{T}),
\end{equation}
where $V_{4}$ is the effective potential in the  full four-dimensional
theory and $V_{3}$ is the effective potential in the three-dimensional
theory. This implies that
\begin{equation}
V_{s}(\sigma)=\frac{1}{2}m^{2}\sigma
-\frac{10\pi^{2}}{3}\frac{\sigma^{3}}{T^{2}}
\end{equation}
and
\begin{equation}
V_{b}(\sigma)=\frac{1}{2}m^{2}\sigma +\frac{\lambda}{4}\sigma^{2}
-\frac{T}{6\pi}(e\sigma)^{\frac{3}{2}}-\frac{T}{12\pi}(2\lambda\sigma)
^{\frac{3}{2}}-\frac{T}{12\pi}(\frac{e^{2}T^{2}}{2}+e^{2}\sigma)^{\frac{3}{2}}.
\end{equation}
{\it The ring improved effective potential.}$\,\,$
The ordinary ring improved effective potential for the chiral Abelian Higgs
model computed in Landau gauge is \cite{arnold1}
\begin{equation}
\label{ring}
V_{ring}(\phi_{0})=\frac{1}{24}(4\lambda +
3e^{2}+g^{2})(T^{2}-T^{2}_{b})\phi_{0}^{2}
-\frac{T}{12\pi}(2M_{T}^{3}+M_{L}^{3}+m_{1}^{3}+m_{2}^{3})
+\frac{1}{4}\lambda\phi_{0}^{4},
\end{equation}
where $T_{b}$ is the barrier temperature:
\begin{equation}
T^{2}_{b}=\frac{12c^{2}}{4\lambda +3e^{2}+g^{2}}
\end{equation}
and the thermal masses are
\begin{eqnarray}
M_{T}^{2}&=&e^{2}\phi_{0}^{2}\\
M_{L}^{2}&=&e^{2}\phi_{0}^{2}+\frac{e^{2}T^{2}}{2}\\
m^{2}_{1}&=&-c^{2}+\lambda\phi_{0}^{2}+(4\lambda +3e^{2}+g^{2})\frac{T^{2}}{12}
\\
m^{2}_{2}&=&-c^{2}+3\lambda\phi_{0}^{2}+(4\lambda +3e^{2}+g^{2})
\frac{T^{2}}{12}.
\end{eqnarray}
Here $\phi_{0}$ is the background field, which is always larger than or equal
to zero. The symmetric phase is represented by $\phi_{0}=0$, while the broken
phase is represented by $\phi_{0}>0$. Notice also that the masses $m_{1}^{2}$
and $m_{2}^{2}$ becomes negative below the barrier temperature, implying
that the effective potential becomes complex.

In fig.~\ref{comp} we have shown $V(\sigma)$ at the critical temperature for a
Higgs mass of 55 GeV, and gauge and Yukawa couplings of 0.45 and 0.6,
respectively. In fig.~\ref{ring2} we have shown the corresponding ring
improved
effective potential. Both potentials show that the symmetry is restored via
a first order phase transition. This is to be expected from a renormalization
group argument, namely that the renormalization group equations do not have a
non-trivial fixed point \cite{gins}.
We alsosee that the form of the potentials in the symmetric
phase is qualitatively the same, although the barrier height is approximately
30$\%$ higher for $V(\sigma)$. In turn, this will affect the phase transition.
Moreover, the potential in the symmetric phase increases rapidly for small
values of $\sigma$.
\section{Summary and Final Remarks}
We have calculated a three-dimensional effective theory for the static modes
in the chiral Abelian Higgs model by integrating out the heavy modes.
The thermal masses are seen to be correctly reproduced. Using this
three-dimensional model, a one-loop calculation for the composite operator
$\phi^{\dagger}\phi$ has been performed.  We have then used the
obtained potential to investigate the phase transition. The potential is
similar to the ring improved effective potential at the critical temperature
and the symmetry is restored via a first order phase transition.

We have also noted that the effective potential is gauge parameter independent
in the one-loop approximation.
The questions of gauge invariance and gauge fixing dependence are
important and will be subject of further
investigation, particularly in connection with the gauge invariant
Vilkovisky-DeWitt effective action \cite{vilko}. This work is in progress.
Finally, it would be of interest to extend the present work
by doing a Hubbard-Stratonovich transformation \cite{hub}.
One could then carry out a two-variable saddle point approximation
for the auxillary fields. This method has previously been applied to
$\lambda\phi^{4}$ and has correctly reproduced the second order phase
transition this model undergoes \cite{skalar}.\\ \\
The author would like to thank Finn Ravndal for useful comments and
suggestions.

\newpage\pagebreak
\setcounter{figure}{0}
\begin{figure}[b]
\underline{FIGURE CAPTIONS:}
\caption{Leading contributions to the scalar mass in the high temperature
limit.}
\label{fey1}
\caption{Dominant contribution to the vector mass in the high temperature
limit.}
\label{fey2}
\caption{Effective potential for the composite operator $\sigma
=\phi^{\dagger}\phi$ at the critical
temperature. The Higgs mass is 55 GeV, $g=0.6$ and  $e=0.45$.}
\label{comp}
\caption{Ring improved effective potential at the critical temperature.
The Higgs mass is 55 GeV, $g=0.6$ and $e=0.45$.}
\label{ring2}
\end{figure}
\end{document}